\documentclass[%
twocolumn,
showpacs,
nofootinbib,
nobibnotes,
amsmath,amssymb,
aps,
pra,
floatfix,
]{revtex4}
\usepackage{graphicx}
\usepackage{dcolumn}
\usepackage{bm}
\usepackage{hyperref}
\usepackage{natbib}
\usepackage{subfigure}
\usepackage{color}

\newcommand{\tr}{\mathrm{tr}}
\begin{document}

\title{Einstein-Podolsky-Rosen steering in\\ non-Hermitian, nonlinear soliton surface plasmons}
\author{Ali \"{U}. C. Hardal}
\email{ahardal@ku.edu.tr}
\affiliation{Department of Physics, Ko\c{c} University, \.Istanbul, 34450, Turkey}
\begin{abstract}
We investigate the dynamics of Einstein-Podolsky-Rosen steering in a soliplasmonic system where the asymmetrical coupling between linear solitons and nonlinear plasmons is controlled by the weak soliton amplitude. We employ non-Hermitian quantum dynamics to evaluate quantum steering witnesses for different coupling strengths. Quantum steering is examined for an initially Fock state and we briefly discuss Einstein-Podolsky-Rosen entanglement for initially coherent states. We numerically investigate the adjustability of the witnessing periods of quantum steering via tunable coupling strength of soliton-plasmon interaction and discuss the possibility of the nonlinear soliplasmonic systems as a robust platform for implementing one-sided quantum key distributions.
\end{abstract}
\maketitle
\section{Introduction}\label{sec:intro}
Non-Hermitian quantum interactions~\cite{bender1998real,bender1999pt,mostafazadeh2002pseudo,mostafazadeh2003exact,bender2007making} have been recognised in hybridised metamaterials~\cite{PhysRevA.87.053824}, coupled microcavities~\cite{peng2014parity}, waveguides~\cite{PhysRevLett.101.080402,PhysRevA.87.013816}, optical lattices~\cite{PhysRevLett.110.223902, longhi2014} and Bose-Einstein condensates~\cite{graefe2008non,graefe2008mean,dast2013eigenvalue}. Emergence of many intriguing implementations from non-Hertmiticity such as PT-symmetric lasers~\cite{PhysRevLett.105.053901,PhysRevLett.108.173901,PhysRevA.84.063833} and invisibility~\cite{PhysRevA.82.043803,PhysRevA.82.031801,mostafazadeh2013invisibility,PhysRevLett.106.213901,lg15in} have also been reported.

While a generic asymmetrically coupled quantum system can reveal many interesting phenomena~\cite{santos2012non,karakaya,santos2014non,hardal2014dynamics}, here, we consider a nonlinear soliton-plasmon system that is characteristically non-Hermitian~\cite{milian2012soliton,ferrando2013variational,karakaya}. The model system consists of a linear (soliton) and a non- linear (plasmon) waveguides coupled via intensity dependent particle exchange controlled by the weak soliton amplitude. 

Our first objective is to investigate the non-Hermitian dynamics of the Einstein-Podolsky-Rosen (EPR) steering~\cite{einstein1935can, schrodinger1935discussion} and then show the possibility of manipulation of steering via controlling the asymmetric coupling strength. The end result leads us to recognise the nonlinear soliplasmon system as a possible candidate for implementing one-sided quantum key distributions (1SQKD)~\cite{branciard2012one, bennett1992quantum}. 

In this contribution, EPR steering is witnessed via a set of inequalities~\cite{cavalcanti2011unified,he2012einstein,olsen2015asymmetric} by employing non-Hermitian quantum dynamics~\cite{graefe2008mean,sergi2013non,dattoli1990non,brody2012mixed}. We consider an initial state in which the soliton mode is in a single photon Fock state and the plasmon mode is in its vacuum. While the previous works reported the existence of close relations between coherence and correlations as well as the population dynamics in non-Hermitian coupled cavity systems~\cite{karakaya,hardal2014dynamics}, we found that quantum steering is absent for such a initial preparation (see also Ref.~\cite{olsen2015asymmetric}). Nevertheless, we provide discussions of EPR entanglement if the system is prepared in an initially coherent state, as well. Finally, controllable interaction strength between soliton and surface plasmons via soliton amplitude enable us to extend the witnessing period of EPR steering. We exploit this unique property of asymmetrical coupling and discuss the robustness of one-way quantum steering in non-linear soliplasmonic systems. ~\cite{bowles2014one,branciard2012one}.

Experimental studies on EPR steering can be found in Refs.~\cite{schneeloch2013violation,steinlechner2013strong,sun2014experimental,wittmann2012loophole,armstrong2015multipartite,bennet2012arbitrarily}. In particular, one-way quantum steering has been demonstrated with Gaussian beams~\cite{handchen2012observation} as well as with entangled photon pairs very recently~\cite{wollmann2015observation}. On the other side, experiments with plasmonic matematerials concentrated on quantum state transfer~\cite{wang2012hong, zhou2012characterizing}, quantum interference~\cite{jha2015metasurface}, coherent absorption of photons~\cite{roger2015coherent} and distillation of quantum entanglement~\cite{asano2015distillation}. For recent advances in theoretical and experimental quantum plasmonics, we refer to Ref.~\cite{tame2013quantum}.

This paper is organised as follows. In Section~\ref{sec:model}, we introduce the model system and briefly discussed the non-Hermitian quantum dynamics. In Section~\ref{sec:results}, we first provide the definitions of steering witnesses to be used and  then continue with our results for the dynamics of EPR steering in Section~\ref{sec:epr_1}. In Section~\ref{sec:epr_2}, we discuss the robustness and the controllability of one-way steering via asymmetrical coupling strength. We conclude in Section~\ref{sec:conc}.
\section{The Model}\label{sec:model}
We consider a system in which spatial solitons in a nonlinear Kerr medium resonantly interact with surface plasmons in a metal through a dielectric layer~\cite{milian2012soliton,ferrando2013variational}.
The system can be considered as a coupled linear (soliton) and nonlinear (plasmon) waveguides where the coupling is controlled by the weak soliton amplitude. The dynamics
is governed by a Hamiltonian of the form~\cite{karakaya}
\begin{eqnarray}\label{eq:soliplasmon}
\nonumber \hat{\mathbf{H}}&=&\omega\hat{\mathbf{a}}^{\dagger}\hat{\mathbf{a}}+\omega\hat{\mathbf{b}}^{\dagger}\hat{\mathbf{b}}+U\hat{\mathbf{a}}^{\dagger}\hat{\mathbf{a}}^{\dagger}\hat{\mathbf{a}}\hat{\mathbf{a}}\\
&+&g_{\mathrm{a}\mathrm{b}}\hat{\mathbf{a}}\hat{\mathbf{b}}^{\dagger}+g_{\mathrm{b}\mathrm{a}}\sqrt{\hat{\mathbf{n}}_{\mathrm{a}}}\hat{\mathbf{a}}^{\dagger}\hat{\mathbf{b}},
\end{eqnarray}
where $\hat{\mathbf{a}}$~($\hat{\mathbf{a}}^{\dagger}$) and $\hat{\mathbf{b}}$~($\hat{\mathbf{b}}^{\dagger}$) are the annihilation (creation) operators for the soliton and surface plasmon modes, respectively. $\omega$ is the resonance frequency for the field modes and $U$ quantifies the strength of the nonlinear Kerr interaction. We denote the linear soliton-plasmon coupling strength with $g_{\mathrm{a}\mathrm{b}}$, while the amplitude dependent, nonlinear coupling strength is determined via $g_{\mathrm{b}\mathrm{a}}\sqrt{\hat{\mathbf{n}}_{\mathrm{a}}}$ with $\hat{\mathbf{n}}_{\mathrm{a}}:=\hat{\mathbf{a}}^{\dagger}\hat{\mathbf{a}}$.

As it is easy to spot, the system Hamiltonian~(\ref{eq:soliplasmon}) is characteristically non-Hermitian even if the coupling strengths are equal, i.e., $g_{\mathrm{a}\mathrm{b}}=g_{\mathrm{b}\mathrm{a}}$. Therefore, we employ a general formalism~\cite{graefe2008mean,sergi2013non,dattoli1990non,brody2012mixed} and write the system Hamiltonian~(\ref{eq:soliplasmon}) as the sum of its Hermitian  $\hat{\mathbf{H}}_+$ and anti-Hermitian $\hat{\mathbf{H}}_-$ parts: $\hat{\mathbf{H}}=\hat{\mathbf{H}}_++\hat{\mathbf{H}}_-$, $\hat{\mathbf{H}}_{\pm}:=(1/2)(\hat{\mathbf{H}}\pm\hat{\mathbf{H}}^{\dagger})$ and $\hat{\mathbf{H}}_{\pm}=\pm\hat{\mathbf{H}}_{\pm}^{\dagger}$. Hence, the evolution of any state $\hat{\mathbf{\rho}}(t)$ of the system is evaluated by the master equation
\begin{equation}\label{eq:master}
\frac{\partial}{\partial t}\hat{\rho}(t)=-i[\hat{\mathbf{H}}_+,\hat{\rho}(t)]_+-i[\hat{\mathbf{H}}_-,\hat{\rho}(t)]_-,
\end{equation}
where $[,]_-$ and $[,]_+$ denote the commutator and the anti-commutator of the corresponding operators. Due to the non-unitary nature of the master equation~(\ref{eq:master}), the normalization is guaranteed via the mapping $\hat{\rho}(t)\rightarrow [\hat{\rho}(t)/\tr(\hat{\rho}(t))]$ and therefore, the expectation value of a given operator $\hat{\mathbf{Q}}$ is calculated as $\langle\hat{\mathbf{Q}}\rangle=\tr[\hat{\mathbf{Q}}\hat{\rho}(t)]/\tr(\hat{\rho}(t))$.

Next section shall start with the introduction of the steering witnesses that are going to be calculated. We, then, continue with our results for the dynamics of EPR steering for an initially single photon Fock state. Finally, we discuss the possible usage of the nonlinear soliplasmons as a possible platform for the implementation of one-sided QKDs.
\section{Results}\label{sec:results}
Let us start with the definition of EPR steering witness for an $N$-mode system~\cite{cavalcanti2011unified,he2012einstein}
\begin{equation}\label{eq:str_wtns}
\bigg|\bigg\langle\prod_{k=1}^{N}\hat{\mathbf{a}}_k\bigg\rangle\bigg|^2>\bigg\langle\hat{\mathbf{n}}_1\prod_{k=2}^{N}\bigg(\hat{\mathbf{n}}_k+\frac{1}{2}\bigg)\bigg\rangle,
\end{equation}
where $\hat{\mathbf{n}}_k:=\hat{\mathbf{a}}_k^{\dagger}\hat{\mathbf{a}}_k$. In a two-mode system, since the either modes can steer the other, it is rather more convenient to define~\cite{olsen2015asymmetric}
\begin{eqnarray}\label{eq:strs}
\label{eq:strs1}\zeta_{\mathrm{a}\mathrm{b}}&:=&\langle\hat{\mathbf{a}}^{\dagger}\hat{\mathbf{b}}\rangle\langle\hat{\mathbf{b}}^{\dagger}\hat{\mathbf{a}}\rangle-\langle\hat{\mathbf{a}}^{\dagger}\hat{\mathbf{a}}(\hat{\mathbf{b}}^{\dagger}\hat{\mathbf{b}}+1/2)\rangle,\\
\label{eq:strs2}\zeta_{\mathrm{b}\mathrm{a}}&:=&\langle\hat{\mathbf{a}}^{\dagger}\hat{\mathbf{b}}\rangle\langle\hat{\mathbf{b}}^{\dagger}\hat{\mathbf{a}}\rangle-\langle\hat{\mathbf{b}}^{\dagger}\hat{\mathbf{b}}(\hat{\mathbf{a}}^{\dagger}\hat{\mathbf{a}}+1/2)\rangle,
\end{eqnarray}
with $\hat{\mathbf{a}}_1:=\hat{\mathbf{a}}$ and $\hat{\mathbf{a}}_2:=\hat{\mathbf{b}}$. If $\zeta_{\mathrm{a}\mathrm{b}}>0$, we say the mode $\hat{\mathbf{a}}$ steers the mode $\hat{\mathbf{b}}$. Equivalently, if $\zeta_{\mathrm{b}\mathrm{a}}>0$, we say the mode $\hat{\mathbf{b}}$ steers the mode $\hat{\mathbf{a}}$. 
\subsection{EPR steering}\label{sec:epr_1}
We consider an initial state $|\psi(0)\rangle$ such that the soliton mode is in a single photon Fock state while the plasmon mode is in its vacuum
\begin{equation}\label{eq:initial_state}
|\psi(0)\rangle=|1\rangle_{\mathrm{a}}|0\rangle_{\mathrm{b}}.
\end{equation}
Throughout our numerical analysis, we set $U=-0.01$, $g_{\mathrm{a}\mathrm{b}}=\kappa g$, $g_{\mathrm{b}\mathrm{a}}=g$ with $g=0.1$ (all the parameters are scaled with the resonance frequency $\omega$). We examine three cases for which $\kappa=1$, $\kappa=2$ and $\kappa=0.5$, respectively. 
\begin{figure}[!t]
\centering
     \begin{center}
        \subfigure[]{
            \label{fig:fig1a}
            \includegraphics[width=8.25cm]{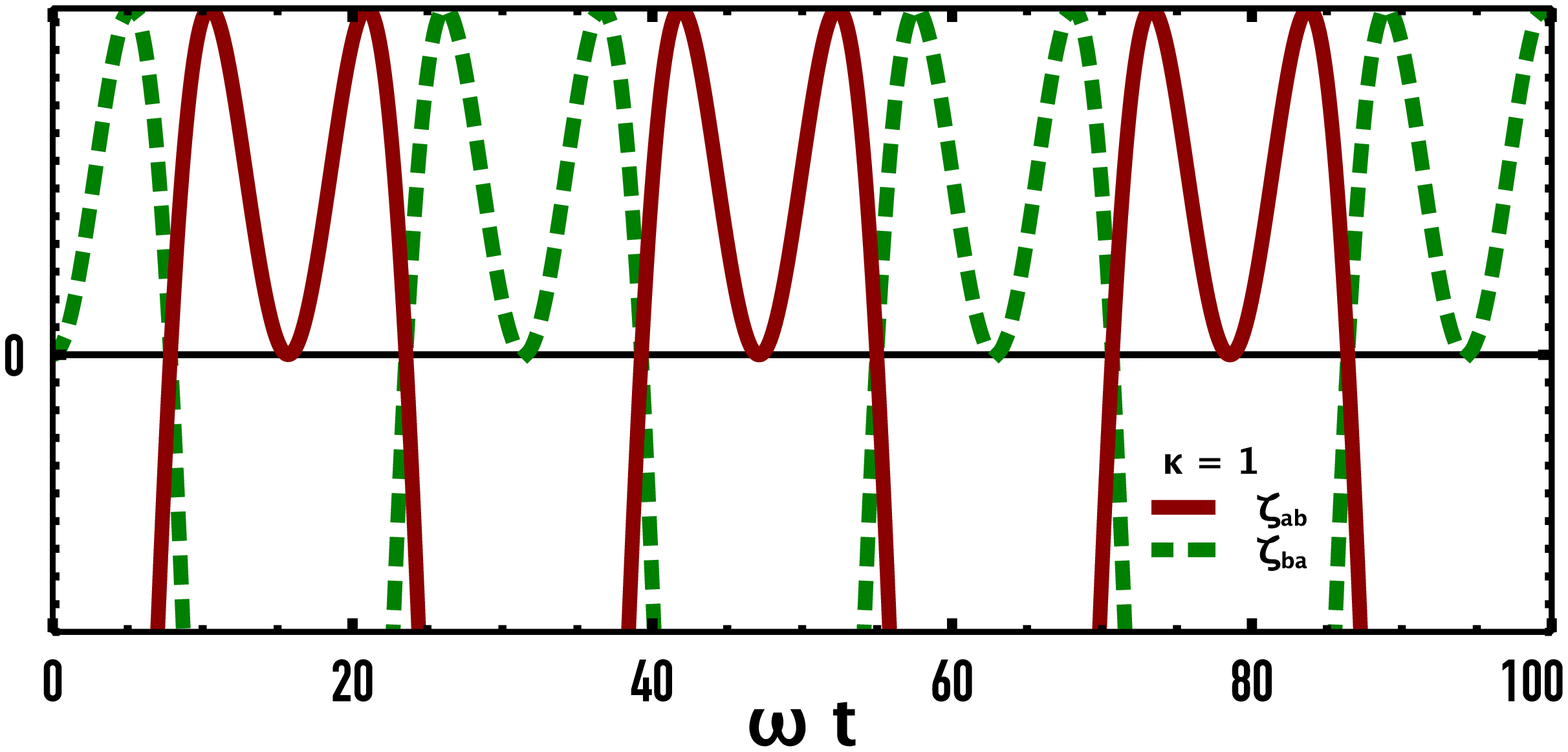}
        }\\
        \subfigure[]{
           \label{fig:fig1b}
           \includegraphics[width=8.25cm]{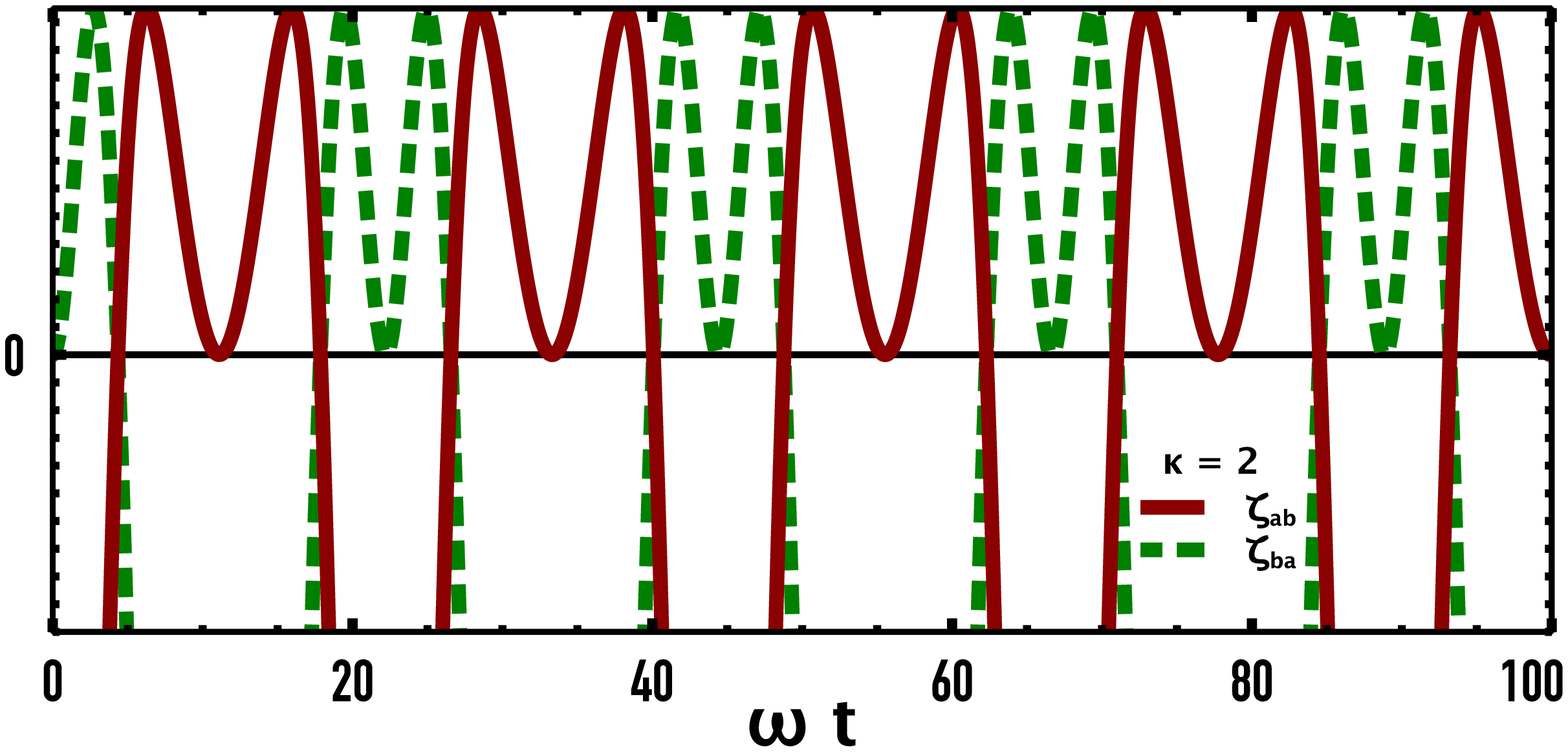}
        }\\
         \subfigure[]{
           \label{fig:fig1c}
           \includegraphics[width=8.25cm]{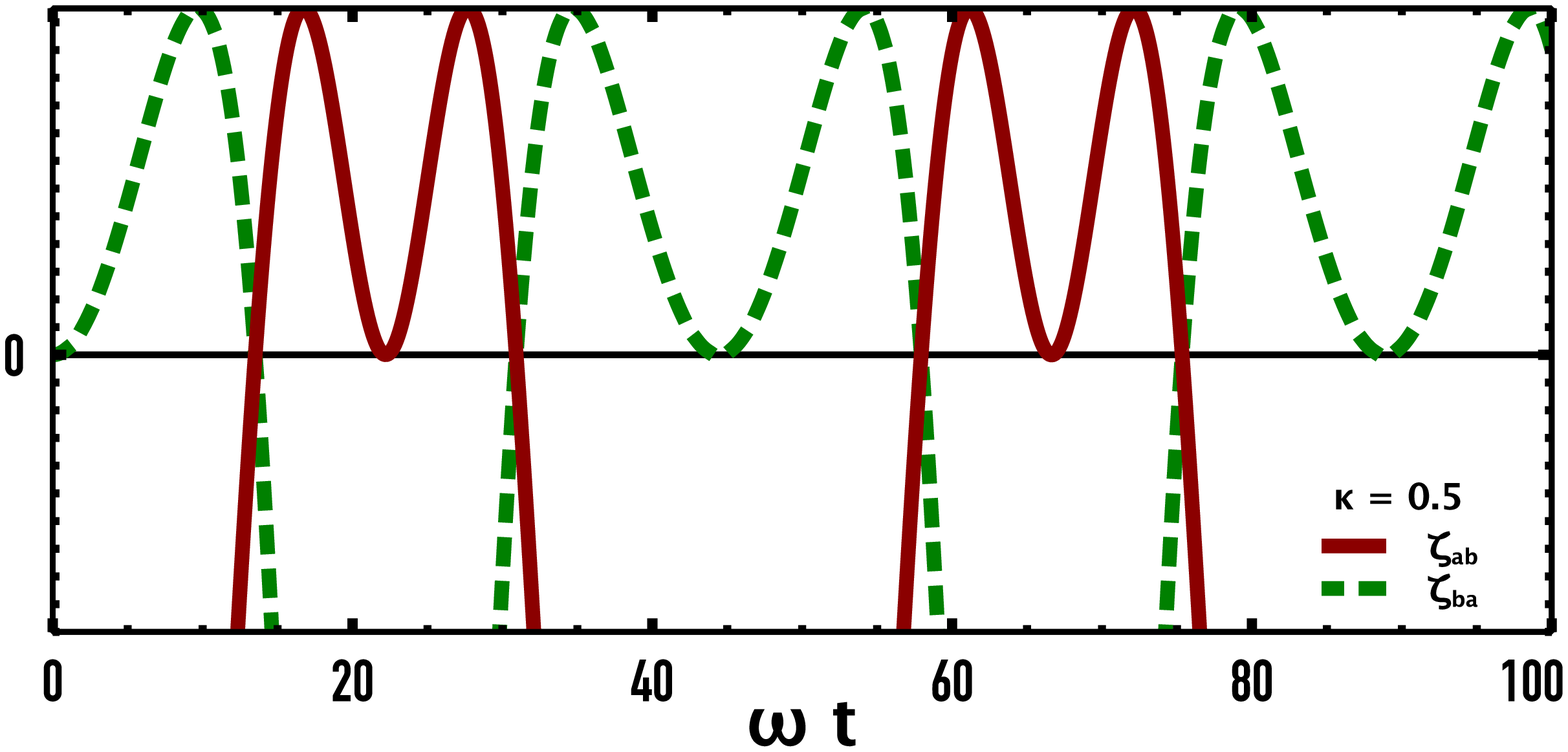}
        }

   \end{center}
    \caption{Dynamics of steering parameters $\zeta_{\mathrm{a}\mathrm{b}}$ (dark-red) and $\zeta_{\mathrm{b}\mathrm{a}}$ (green, dashed) with respect to scaled time $\omega t$ for a) $\kappa=1$, b) $\kappa=2$ and c)$\kappa=0.5$.
    All the other parameters are as explained in the text.
                }
   \label{fig:fig1}
\end{figure}

Our typical results are shown in Fig.~\ref{fig:fig1}. Before proceed and discuss the results, let us identify time intervals $T_{ab}$ and $T_{ba}$ therein the calculated values of the steering witnesses becomes greater than zero, reach their maxima and finally becomes zero again. We simply call $T_{ab}$ and $T_{ba}$ witnessing periods.

For $\kappa=1$, we have $T_{ab}=T_{ba}$ and steering is asymmetrical, i.e., $\zeta_{ab}$ and $\zeta_{ba}$ are never different than zero at the same time. The asymmetric behaviour with equal witnessing periods agrees with the previous investigations of quantum steering in coupled cavity arrays~\cite{olsen2015asymmetric}. 

The results of the broken exchange symmetry due to the settings where $\kappa=2$ or $\kappa=0.5$ are depicted in Figs~\ref{fig:fig1b} and~\ref{fig:fig1c}, respectively. $T_{ab}$ stays (almost completely) unaffected, while the witnessing period $T_{ba}$ behaves as a function that is inversely proportional with the parameter $\kappa$, i.e., $T_{ba}\propto(1/\kappa)$. Such inverse proportionality to the coupling strength of initially Fock states has also been reported for genuine mode correlations in generic asymmetrically coupled cavity systems~\cite{hardal2014dynamics}. 

The difference of the reactions of the periods $T_{ab}$ and $T_{ba}$ to the change of the exchange symmetry is solely due to the choice of the initial state and can be reversed by choosing $|\psi(0)\rangle=|0\rangle_{\mathrm{a}}|1\rangle_{\mathrm{b}}$.

We finalize this section by noting that our simulations with an initially coherent state, $|\psi(0)\rangle=|\alpha\rangle_{\mathrm{a}}|0\rangle_{\mathrm{b}}$ with $\alpha=1$, showed no witnessing of quantum steering within the considered time intervals (see also Ref.~\cite{olsen2015asymmetric}). However, up to third order, EPR type entanglement between the modes has been witnessed by using an equivalent formalism proposed in Ref.~\cite{he2012einstein}. We found that while the third order correlations are almost always present within the considered time interval, first order (mode type) correlations are hardest to detect. Second order (particle type) correlations detectable within short time intervals and the total dynamics shows collapse and revivals specific to this order. Increase in the local nonlinearity $U$ favours the second order correlations as expected. The case of strong plasmon excitation, $\kappa=2$, increases the witnessing periods of second and third order correlations (see also Ref.~\cite{karakaya}).
\subsection{A stage for one-sided QKDs}\label{sec:epr_2}
QKDs~\cite{gisin2002quantum, lo2014secure} are quantum cryptographic tasks in which two distinct parties share a secret key to encrypt and decrypt any communication. As usual, let us call our communicators Alice and Bob. The key distribution where both Alice and Bob can trust their measurement instruments commonly referred as the standard QKD. In this case quantum entanglement is sufficient to achieve the end result~\cite{nielsen2010quantum}. If, however, neither of their instruments can be trusted, then Alice and Bob should demonstrate the violation of Bell inequalities to ensure the security of the communications~\cite{acin2007device}.

The case that we are interested in achieves QKD in such a way that one party, say Alice, cannot trust her instrument, while Bob can. Then, Alice should steer Bob to ensure a secure key distribution. In fact, systems that demonstrate one-sided QKDs should violate the quantum steering inequalities~\cite{cavalcanti2009experimental} (see also Ref.~\cite{wiseman2007steering} for the classification of quantum nonlocality and Ref.~\cite{branciard2012one} for the visualisation
of this classification from perspective of QKDs).  
\begin{figure}
\begin{center}
\includegraphics[width=6.5cm]{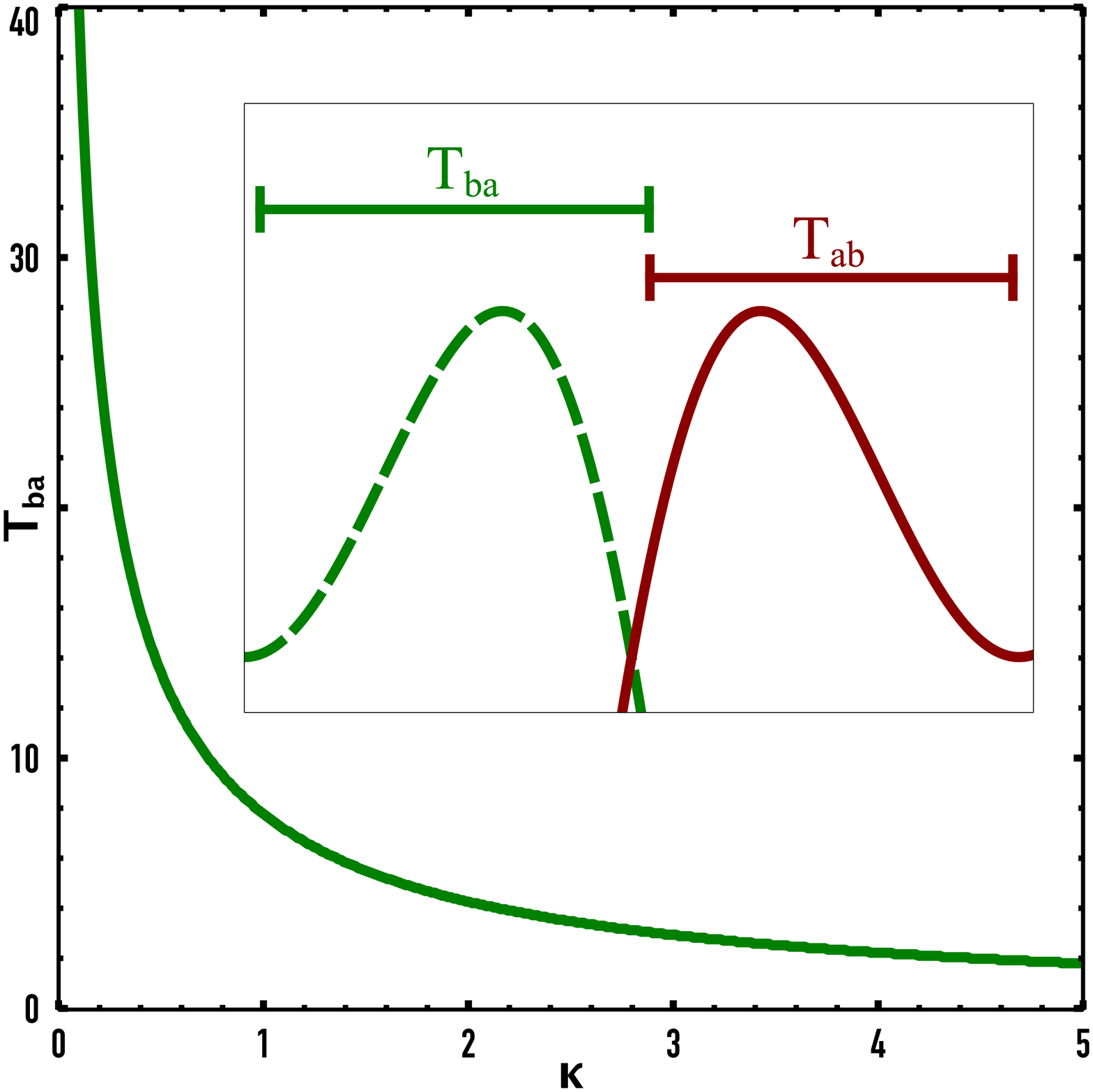}
\caption{The change in the witnessing period $T_{ba}$ of the steering parameter $\zeta_{ba}$ with respect to $\kappa$. All the other parameters are as explained in the text.}
\label{fig:fig2}
\end{center}
\end{figure}

Let us, now, exploit the uneven interactions between solitons and surface plasmons to achieve a robust one-way quantum steering~\cite{bowles2014one,branciard2012one}. In Fig.~\ref{fig:fig2}, we plot the witnessing period $T_{ba}$ with respect to the parameter $\kappa\in[0.1,5]$. At the extremes, $\kappa\ll1$ and $\kappa\gg1$,  our witnesses $\zeta_{ab}$ and $\zeta_{ba}$ stays positive between their two succeeding maxima shown in Fig.~\ref{fig:fig1}. Therefore, within these limits we may also consider
$T^{\prime}_{ab}=2T_{ab}$ and $T^{\prime}_{ba}=2T_{ba}$ due to the mirror symmetry, as well. However, the change in $T_{ab}$ with respect to $\kappa$ is not drastic as in the case of $T_{ba}$. It assumes close values within the considered $\kappa$ range relative to that of $T_{ba}$ (e.g., $T^{\prime}_{ab}\sim16\omega t$ at $\kappa=1$, $T^{\prime}_{ab}\sim18\omega t$ at $\kappa=0.1$ and $T^{\prime}_{ab}\sim24\omega t$ at $\kappa=5$ ). Secondly, it is numerically more efficient and less demanding to investigate, without any lost of physical information, $T_{ba}$ as the period from $\zeta_{ba}=0$ at $\omega t=0$ to $\zeta_{ba}=0$ at $\omega t>0$.

A guided intuition and straightforward numerical investigation reveal that
\begin{equation}\label{tba}
T_{ba}(\kappa)\sim\frac{a}{\kappa}+\frac{b}{3\kappa^3}+\frac{c}{5\kappa^5},
\end{equation}
up to first three terms. Here, $a=8.4543\pm0.0198$, $b=-0.5091\pm0.0102$ and $c=0.0069\pm0.1\times10^{-4}$. For $\kappa\geqslant1$ we recover $T_{ba}\propto(1/\kappa)$. In the limit $\kappa\ll1$, we require higher order terms to be able to have more accurate analytic description.

An intriguing argument can be made via observing $\lim_{\kappa\rightarrow0}T_{ba}=\infty$ (we can be mathematically more rigorous by stating $\lim_{\kappa\rightarrow0^{+}}T_{ba}=\infty$, though it seems unnecessary for our considered parameter regimes). That is to say, for $\kappa\ll1$, we have $T_{ba}\gg T_{ab}$. By taking into account the our choice of initial state that leads $\zeta_{ba}>0$ after immediately $\omega t=0$, we obtain a robust one-way quantum steering. Since we have $\zeta_{ba}>0$ and $\zeta_{ab}<0$ within the period $T_{ba}$, the latter results suggest a controllable, configurable one-sided QKD setup. 

The interplay between non-Hermitian dynamics and coherence leading to the non-conservation of the mean number of excitations as well as the mean spin has already been reported in the literature~\cite{karakaya,hardal2014dynamics}. For an initially Fock state the latter symmetries of the system are conserved during the dynamical evolution. However, it has been verified that the effects of the non-Hermitian dynamics with uneven exchange rates between the modes emerge as the change in periods of the dynamical oscillations ($\sim (1/\kappa)$) of the corresponding observable, even if the system is prepared in an initially Fock state~\cite{hardal2014dynamics}. Since the steering witnesses $\zeta_{ba}$ and $\zeta_{ab}$ are the functions of exchange rates and of the mean number of excitations, the broken symmetry affects the witnessing periods, inherited from the mean number dynamics.  

Lastly, we recall that the parameters $\zeta_{ba}$ and $\zeta_{ab}$ are not measures of quantum correlations but mere witnesses. Therefore, $T_{ba}\gg T_{ab}$ with $\zeta_{ba}>0$ and $\zeta_{ab}<0$ is not necessarily mean the subsystem $\hat{\mathbf{a}}$ does not steer the subsystem $\hat{\mathbf{b}}$. On the other hand, as long as we know, at our end, that we can achieve robust quantum steering for a sufficient period of time to establish a secure communication, the device (method) used by the other party is \textit{less relevant}. 
\section{Conclusions}\label{sec:conc}
In summary, we investigated the non-Hermitian dynamics of EPR steering in a nonlinear soliplamonic system. The dynamics is investigated by solving quantum master equation equipped for non-Hermitian Hamiltonians. Quantum steering is witnessed for an initially Fock state, while a brief discussion on EPR entanglement is given for initially coherent states for which the quantum steering is absent. The intensity dependent, asymmetrical coupling between solitons and surface plasmons enable us to control the witnessing periods of quantum steering. The control of the coupling strength is associated with a parameter flagged as $\kappa$. Quantum steering is examined by setting $\kappa\in[0.5, 1, 2]$ for an initial profiling. With the possibility of an application for quantum key distributions in our mind, the results obtained for the dynamics of quantum steering after the latter choices of $\kappa$ are guided us to consider even wider intervals of coupling strength. We numerically reveal that for $\kappa\ll1$ one can achieve robust one-way quantum steering with a careful choice of initial preparation for the coupled quantum system .

\end{document}